\documentclass[noacm,sigconf]{acmart}

\usepackage{booktabs} 
\usepackage[utf8]{inputenc}
\usepackage{adjustbox}
\usepackage{paralist}
\usepackage{graphicx}
\usepackage{subcaption}
\usepackage{multirow}
\usepackage{bbm}
\usepackage{amsmath}
\usepackage{amsfonts}
\usepackage{amssymb}
\usepackage{url}
\usepackage{color}
\usepackage{mathtools}
\usepackage{tkz-graph}
\usepackage{tikz}
\usetikzlibrary{backgrounds}
\usetikzlibrary{trees,positioning,arrows}
\usepackage{wrapfig}
\usepackage[ruled,linesnumbered,lined,boxed]{algorithm2e}
\usepackage{array}
\usepackage{bm}
\usepackage{acronym}
\usepackage{enumerate}
\usepackage[shortlabels]{enumitem}

\acrodef{ALS}{Alternating Least Square}
\acrodef{VAE}{Variational Auto-Encoder}
\acrodef{PMF}{Probabilistic Matrix Factorization}
\acrodef{GBS}{Generalized Binary Search}
\acrodef{GP}{Gaussian Process}
\acrodef{SBS}{Sequential Bayesian Search}

\newcommand{\RQ}[2]{
    \begin{description}[topsep=0pt,nosep=0pt, leftmargin=0.75cm, noitemsep,nolistsep]
    \phantomsection\label{section:setup:rq#1}
    \item[RQ#1] #2
    \end{description}
}
\newcommand{\RQRef}[1]{\textbf{\hyperref[section:setup:rq#1]{RQ#1}}}

\newcommand{\rbr}[1]{\left(#1\right)}

\newcommand{\set}[1]{\left\{#1\right\}}
\newcommand{\norm}[1]{\|#1\|}
\newcommand{\diag}[1]{\text{diag}(#1)}

\setcopyright{acmlicensed}
\acmConference[SIGIR '20]{Proceedings of the 43rd International ACM SIGIR Conference on Research and Development in Information Retrieval}{July 25--30, 2020}{Virtual Event, China}
\acmBooktitle{Proceedings of the 43rd International ACM SIGIR Conference on Research and Development in Information Retrieval (SIGIR '20), July 25--30, 2020, Virtual Event, China}

\let\cal\mathcal

\begin{document}
\fancyhead{}

\title{Towards Question-based Recommender Systems}

\author{Jie Zou}
\affiliation{%
  \institution{University of Amsterdam}
  \city{Amsterdam}
 \country{The Netherlands}}
\email{j.zou@uva.nl}

\author{Yifan Chen}
 \affiliation{%
  \institution{National University of Defense Technology}
  \city{Changsha}
  \country{China}}
\email{yfchen@nudt.edu.cn}

  \author{Evangelos Kanoulas}
\affiliation{%
  \institution{University of Amsterdam}
  \city{Amsterdam}
  \country{The Netherlands}}
\email{e.kanoulas@uva.nl}

\renewcommand{\shortauthors}{J. Zou et al.}

\begin{abstract}
Conversational and question-based recommender systems have gained increasing attention in recent years, with users enabled to converse with the system and better control recommendations. Nevertheless, research in the field is still limited, compared to traditional recommender systems. In this work, we propose a novel \textbf{Q}uestion-based \textbf{rec}ommendation method, Qrec, to assist users to find items interactively, by answering automatically constructed and algorithmically chosen questions. Previous conversational recommender systems ask users to express their preferences over items or item facets. Our model, instead, asks users to express their preferences over descriptive item features. The model is first trained offline by a novel matrix factorization algorithm, and then iteratively updates the user and item latent factors online by a closed-form solution based on the user answers. 
Meanwhile, our model infers the underlying user belief and preferences over items to learn an optimal question-asking strategy by using Generalized Binary Search, so as to ask a sequence of questions to the user. 
Our experimental results demonstrate that our proposed matrix factorization model outperforms the traditional Probabilistic Matrix Factorization model. Further, our proposed Qrec model can greatly improve the performance of state-of-the-art baselines, and it is also effective in the case of cold-start user and item recommendations. 
\end{abstract}

%
%



\maketitle

\section{Introduction}

Online shopping on Internet platforms, such as Amazon, and eBay, is increasingly prevalent, and helps customers make better purchase decisions~\cite{zou2019learning}. 
The high demand for online shopping calls for task-oriented conversational agents which can interact with customers helping them find items or services more effectively~\cite{sun2018conversational}.
This greatly stimulates related research on conversational and question-based recommender systems ~\cite{sun2018conversational, zhang2018towards}.

Traditional recommender systems infer user preferences based on their historical behaviors, with the assumption that users have \emph{static} preferences. Unfortunately, user preferences might evolve over time due to internal or external factors~\cite{priyogi2019preference}. Besides, the quality of traditional recommendations suffers greatly due to the \emph{sparsity} of users' historical behaviors~\cite{schafer2007collaborative}.
Even worse, traditional recommendation systems fail to generate recommendations for new users or new items, for which the historical data is entirely missing: the \emph{cold-start} problem~\cite{schafer2007collaborative}.
Compared to the traditional approaches, question-based and conversational recommender systems overcome these issues by placing the \emph{user in the recommendation loop} ~\cite{sun2018conversational, zhang2018towards}. 
By iteratively asking questions and collecting feedback, more accurate recommendations can be generated for the user.

Work on conversational and question-based recommenders~\cite{christakopoulou2016towards, sun2018conversational, li2018towards, zhang2018towards} demonstrates the importance of interactivity.
~\citet{christakopoulou2016towards} presented a recommender system, which elicits user preferences over items. 
~\citet{sun2018conversational} and ~\citet{li2018towards} train their models on a large number of natural language conversations, either on the basis of predefined and well-structured facets ~\cite{sun2018conversational} or based on free-style dialogues but require dialogues to
mention items ~\cite{li2018towards}. 
~\citet{zhang2018towards} proposed a unified paradigm for product search and recommendation, which constructs questions on extracted item aspects, and utilizes user reviews to extract values as simulated user answers.
While these works have developed a successful direction towards conversational recommendation, research in the field is still limited. ~\citet{christakopoulou2016towards} collects user preferences over items, which is inefficient when the item pool is large and continuously updated.
~\citet{sun2018conversational}, ~\citet{zhang2018towards} and ~\citet{li2018towards} make certain assumptions over their input data, most importantly the availability of historical conversational data, or the availability of hierarchical item facets and facet-value pairs. 
In our work, we drop these assumptions: we only hypothesize that items can be discriminated based on textual information associated with them, e.g.\ descriptions and reviews ~\cite{zou2018technology, zou2019learning}. Our model asks questions based on extracted descriptive terms in the related contents, and beliefs are updated based on collaborative filtering, which is one of the most successful technologies in recommender systems ~\cite{he2017neural, schafer2007collaborative}. 

In this work, we propose a novel \textbf{Q}uestion-based \textbf{rec}ommendation method, Qrec, to assist users to find items interactively \footnote{Source code: https://github.com/JieZouIR/Qrec}. 
Our proposed model 
\begin{inparaenum}
\item follows the works by ~\citet{zou2018technology} and ~\citet{zou2019learning}, and generates questions over extracted informative terms; a question pool is constructed by entities (informative terms) extracted from the item descriptions and reviews.
\item proposes a novel matrix factorization method to initialize the user and item latent factors offline by using user-item ratings; 
\item develops a belief-updating method to track the user's belief (preferences over items), and uses \ac{GBS} ~\cite{nowak2008generalized} 
to select a sequence of questions based on the tracked user belief, aiming at learning to ask discriminative questions to gain new information about the user;
\item asks questions, receives answers, updates the user and item latent factors online accordingly by incorporating feedback from the user based on our proposed matrix factorization algorithm, and also renews the user belief to select the next question to ask.
\item generates a recommendation list based on the final user and item latent factors.
\end{inparaenum}

Our model combines the advantages of collaborative filtering based on matrix factorization and content analysis by querying users about extracted informative terms. The matrix factorization model is able to utilize the rating data and discover \emph{latent} correlation between items, while incorporating question-answering over content information, provides \emph{explicit} content discrimination to assist the recommender systems. 
By iteratively asking questions over informative terms and collecting the immediate feedback from the user, our question-based recommender can \emph{track the shifted user preferences}, clarify the user needs, and improve capturing the true underlying user latent factors and item latent factors. Besides, the information gathered from the user constitutes the new observations to \emph{overcome the sparsity and cold-start problem}.

The main contribution of this paper is three-fold: 
\begin{inparaenum}
\item We propose a novel question-based recommendation method, Qrec, that interacts with users by soliciting their preferences on descriptive item characteristics.
\item We propose a novel framework, that incorporates the online matrix factorization and online users' belief tracking for sequential question asking. 
\item We propose a novel matrix factorization method which can incorporate the offline training and efficient online updating of the user and item latent factors.
\end{inparaenum} 

To the best of our knowledge, this is the first work that incorporates online matrix factorization and question asking for item related features. The evaluation results show that our Qrec model achieves the highest performance compared to state-of-the-art baselines and our model is effective in both user and item cold-start recommendation scenarios. 

\section{Related Work}
\label{sec:relwork}
Recommender systems can be classified into three categories: content-based ~\cite{pazzani2007content}, collaborative filtering ~\cite{kawale2015efficient, he2017neural}, and hybrid~\cite{zhao2016predictive} systems. 
Conversational and question-based recommender systems can extend recommender systems in any of the three categories. Early related attempts include the work by ~\citet{bridge2002towards, carenini2003towards, Mahmood:2007, thompson2004personalized, mahmood2009improving, felfernig2011developing}. 
More recently, different ways of feedback are introduced ~\cite{zhao2013interactive, loepp2014choice, graus2015improving, christakopoulou2016towards, yu2019visual, sardella2019approach, jin2019musicbot, wu2019deep}.
~\citet{zhao2013interactive} studied the problem of interactive collaborative filtering, and proposed methods to extend \ac{PMF} ~\cite{mnih2008probabilistic} using linear bandits to select the item to ask feedback for and incorporate the rating back to the \ac{PMF} output. 
~\citet{loepp2014choice} focused on set-based feedback, while ~\citet{graus2015improving} focused on choice-based feedback to learn the latent factors and perform interactive preference elicitation online. Contrary to these works that update the individual user's latent representation, ~\citet{christakopoulou2016towards} proposed a method to update all user and item latent factor parameters of a PMF variant at every feedback cycle, obtaining absolute and pairwise user feedback on items. 
We refer the reader to ~\citet{he2016interactive} and ~\citet{jugovac2017interacting} for a literature review of interactive recommendation. 
Compared with ~\citet{christakopoulou2016towards}, our model also updates all user and item latent factor parameters but based on our own matrix factorization model. Further, while ~\citet{christakopoulou2016towards} elicit user ratings on items, our Qrec model asks questions about extracted descriptive terms of the items, and learns a strategy of asking sequential questions. Furthermore, the selection of questions in Qrec is adaptive to the change of user preferences, instead of relying on the distribution of the items~\cite{christakopoulou2016towards}. 
Last, ~\citet{christakopoulou2016towards} focus on rating prediction while our work focus on the top-N recommendation. They use semi-synthetic data for which they need to obtain the ground truth of the user's preference to every item (like/dislike) using bootstrapping, and thus simulate user's answers for each question, which is not available in our case.
\begin{figure}
\captionsetup{font={small}}
  \includegraphics[width=1\columnwidth]{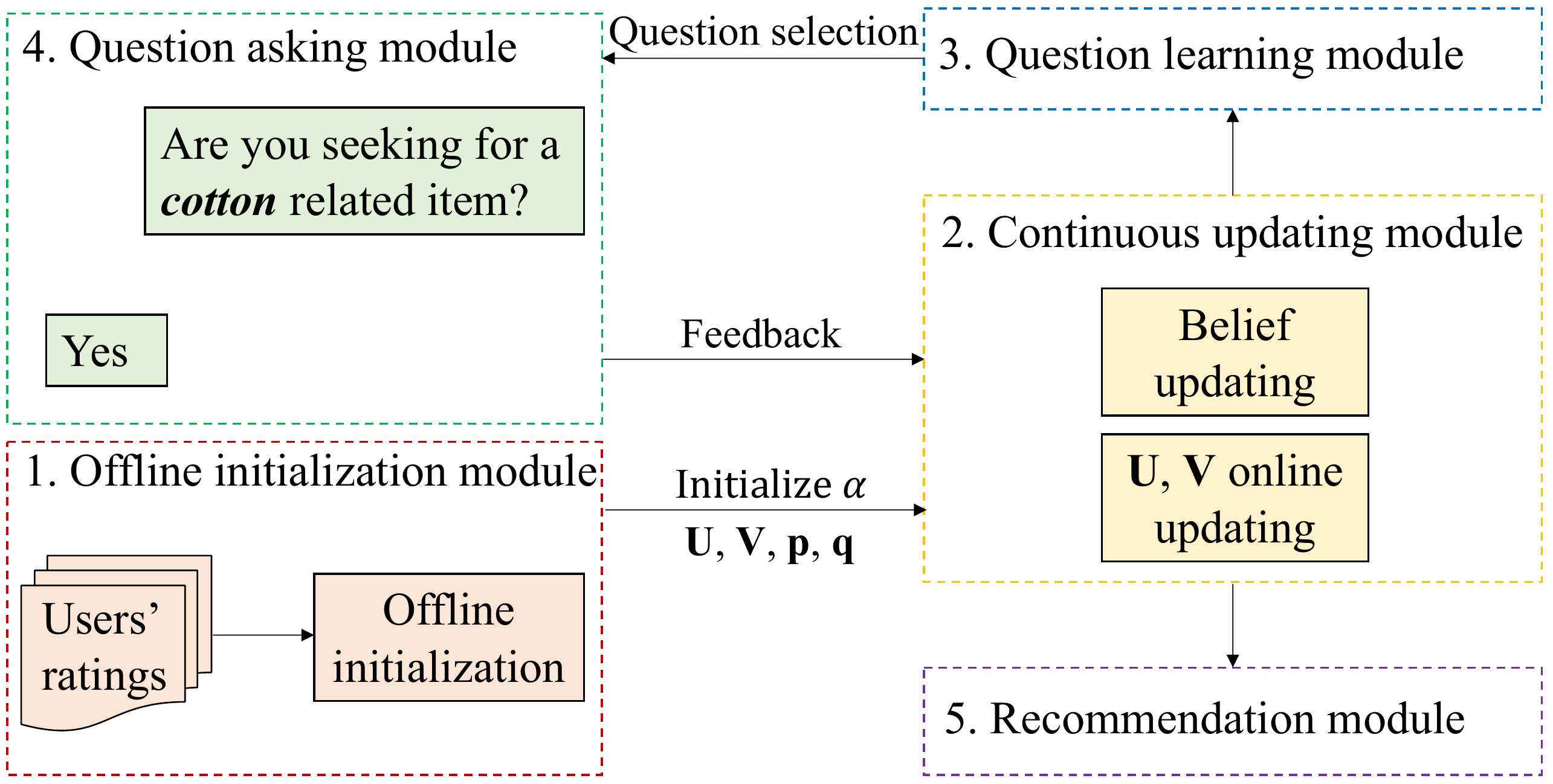}
  \caption{Framework of our proposed question-based recommendation model, Qrec. \textit{Cotton} is an extracted entity (informative term), $\mathbf{U}, \mathbf{V}, \mathbf{p}, \mathbf{q}$ are model variables, and $\alpha$ is a hyper-parameter of user belief.} 
  \label{fig:framework}
\end{figure}

~\citet{zhang2018towards} designed a unified framework for product search and recommendation, and proposed a Personalized Multi-Memory Network (PMMN) architecture for conversational search and recommendation by asking questions over ``aspects'' extracted from user reviews by the means of sentiment labeling. Their model obtains the opinion of the user (i.e.\ value of the aspect-value pair) for the ``aspect'' as feedback. They utilize the user query as an initial query and use the aspect-value pairs of the conversation to expand the representation of the user's query, and thus to match the search and recommendation results. 
Different from this work which uses only the content of user reviews, we incorporate user ratings by collaborative filtering based on our proposed matrix factorization model. Besides, their work trains a model using the data for each user while our online question answering can work without these training data for cold start users and items. Moreover, they query the aspect-value pairs extracted from user review and choose questions based on the log-likelihood of probability estimation over aspects, while we ask questions about descriptive terms of items and select questions based on the user belief tracking and \ac{GBS}.

Reinforcement learning and deep learning on dialogue agents have also been studied for recommendations ~\cite{greco2017converse, christakopoulou2018q, liao2019deep, chen2019towards, lei20estimation}. 
~\citet{sun2018conversational} proposed a deep reinforcement learning framework to build a conversational recommendation agent, which queries users on item facets and focuses on the long-term utility of success or conversion rate. 
~\citet{li2018towards} presented a publicly available dataset called ReDial, and explored a neural method based on dialogue for composing conversational recommendations. They try to predict user opinion over the mentioned items based on the dialogue and sentiment classification to generate a recommendation. On the basis of the ReDial dataset, ~\citet{chen2019towards} proposed a knowledge-based recommender dialog system framework, which incorporates a recommender into a dialog system by using knowledge graphs and transformers. All the aforementioned works are trained on usage data (i.e.\ existing natural language conversations or interactions with the recommender system). 
~\citet{sun2018conversational} require a large number of repeated interactions between the users and the information seeking system to train upon, while ~\citet{li2018towards} and ~\citet{chen2019towards} require mentioning items during the natural language dialogue. Such kind of data is not always available. 
Different from these works, our method does not require such data with large numbers of repeated interactions and mentioned items. 

Learning to ask is another recent and related field of study ~\cite{hu2018playing, wang2018learning}. 
 ~\citet{hu2018playing} presented a policy-based reinforcement learning method to identify the optimal strategy of question selection by continuously learning the probability distribution over all the objects on a 20 Questions game setup. They regard the learned probability distribution on confidence as a state and select the next question according to this state. 
Different from our work, their work introduces data-hungry techniques, which require having large numbers of labeled data and repeated interactions from multiple users for a target item to train upon.
A recent line of work that also involves learning to ask is the work in dialogue and information seeking conversational systems ~\cite{chen2018learning, aliannejadi2019asking}. For example, ~\citet{wang2018learning} studied how to ask good questions in large-scale, open-domain conversational systems with neural question generation mechanisms.
These models need to be trained on existing natural language conversations, which is different from our setup that depends on user ratings. 
 ~\citet{zou2019learning} proposed an interactive sequential Bayesian model for product search. They learn to ask a good question by a cross-user duet training, which learns a belief over product relevance and the rewards over question performances. Different from their work which focuses on a sequential Bayesian product search model based on a cross-user duet training, our model incorporates the user feedback into a matrix factorization model for the recommendation. Further, they require the question answering history and purchase behavior from the same input query for their duet training, while our model does not require having such data.

\section{Methodology}
\label{sec:meth}
In this section, we discuss how we build our question-based recommender system. Our framework shown in Figure~\ref{fig:framework} comprises of five modules: (1) an offline initialization module (Section~\ref{sec:LatentFactor}); (2) a continuous updating module (Section~\ref{sec:LatentFactor}); (3) a question learning module (Section~\ref{sec:Questionlearning}); (4) a question asking module (Section~\ref{sec:Questionasking}); and (5) a recommendation module (Section~\ref{sec:Rec}). 

\subsection{Latent Factor Recommendation}
\label{sec:LatentFactor}
In this section, we describe two of the subcomponents of our Qrec model (shown in Figure~\ref{fig:framework}): the offline initialization module and the continuous updating module. 

Let $\mathbf{R} \in \mathbb{R}^{N \times M}$ be a user-item matrix, and $\mathbf{R}_{i.}$ represents the $i$-th row of $\mathbf{R}$, $\mathbf{R}_{.j}$ represents the $j$-th column of $\mathbf{R}$. Here $N$ and $M$ are the number of users and the number of items, respectively. Similarly, we use $\mathbf{Y}_{i.}$ to represent the $i$-th row of our online affinity matrix $\mathbf{Y} \in \mathbb{R}^{N \times M}$, which is for incorporating user feedback (will be discussed later), use $\mathbf{Y}_{.j}$ to represent the $j$-th column of $\mathbf{Y}$. $\mathbf{U} = [\mathbf{u_1}, \mathbf{u_2}, \ldots, \mathbf{u_i}, \ldots, \mathbf{u_N}]$, $\mathbf{V} = [\mathbf{v_1}, \mathbf{v_2}, \ldots, \mathbf{v_j}, \ldots, \mathbf{v_M}]$, where $\mathbf{u_i}$, $\mathbf{v_j}$ are user and item latent factors respectively. $\mathbf{u_i}$ and $\mathbf{v_j}$ are column vectors. Unless mentioned otherwise, all the vectors in this paper are column vectors. $\mathcal{D}$ is the item collection represented by item documents (descriptions and reviews).

Matrix factorization recommendation techniques have proven to be powerful tools to perform collaborative filtering in recommender systems~\cite{koren2009matrix}. Assume we have $N$ users and $M$ items, matrix factorization decomposes a partially-observed matrix $\mathbf{R} \in \mathbb{R}^{N \times M}$ into two low-rank matrices, the user latent factors $\mathbf{U} \in \mathbb{R}^{N \times K}$ and the item factors $\mathbf{V} \in \mathbb{R}^{M \times K}$ where $K$ is the dimension of user and item latent factors.
The prediction of the unobserved entries in $\mathbf{R}$ is performed as a matrix completion, i.e.\ $\mathbf{R} \approx \mathbf{UV}^\top$. Matrix factorization-based methods have been proposed and successfully applied to various recommendation tasks~\cite{koren2009matrix, kawale2015efficient, christakopoulou2016towards, mnih2008probabilistic}. In matrix factorization, users and items are mapped to the same latent space. Items that have been co-liked by users will lie close in a low dimensional embedding space (latent vector). 

In this paper, we propose a novel model to perform the matrix factorization recommendation, and we refer to it as QMF. The generative process for our model is:
\begin{enumerate}[1.]
	\item For each user $i=1,\ldots,M$, draw a user latent factor $\mathbf{u_i}\sim\mathcal{N}(0,\lambda_u^{-1}\mathbf{I})$;
	\item For each item $j=1,\ldots,N$, draw an item latent factor $\mathbf{v_j} \sim \mathcal N (0, \lambda_v^{-1}\mathbf{I})$.
	\item For each user-item pair $(i, j)\in \mathbf{R}$, draw $R_{ij} \sim \mathcal{N}(\mathbf{p}^\intercal (\mathbf{u_i}\circ \mathbf{v_j}),1)$.
	\item In each user session targeting at a certain item, for each user-item pair $(i, j')\in \mathbf{Y}$, draw $Y_{ij'}\sim \mathcal{N}(\mathbf{q}^\intercal (\mathbf{u_i}\circ \mathbf{v_{j'}}),\gamma^{-1}\mathbf{I})$ for each question asked.
\end{enumerate}
In the above, $\lambda_u, \lambda_v$ are the hyper-parameters modeling the variances in latent vectors, and $\gamma$ is a hyper-parameters modeling the variance in $Y_{ij'}$. 
$\mathbf{p}$ and $\mathbf{q}$ are the free parameters of column vector with $K$ dimension for $R_{ij}$ and $Y_{ij}$, respectively. The intuition behind is that $\mathbf{p}$ and $\mathbf{q}$ can capture some general information across users and items.

\subsubsection{Optimization}
When optimizing our model, the maximization of posterior distributions over $\mathbf{U}$ and $\mathbf{V}$ can be formulated as follows according to the generative process:
\begin{equation}
\small
\max \limits_{\mathbf{U},\mathbf{V},\mathbf{p},\mathbf{q}} p (\mathbf{U},\mathbf{V},\mathbf{p},\mathbf{q} | \mathbf{R}, \mathbf{Y}, \lambda_u, \lambda_v,  \lambda_p,  \lambda_q, \gamma) .
\end{equation}

\noindent Then the maximization of the posterior probability can be reformulated as the minimization of its negative logarithm, which is

\begin{equation}\label{eq:log-post}
\small
\begin{aligned}
    &&& -\log p(\mathbf{U},\mathbf{V}\mid \mathbf{R},\mathbf{Y},\Theta) \\
    & \propto && \frac{1}{2}\sum_{i,j\in \mathbf{R}} \rbr{R_{ij}-\mathbf{p}^\intercal (\mathbf{u_i}\circ \mathbf{v_j})}^2+\frac{\gamma}{2}\sum_{i,j\in \mathbf{Y}} \rbr{Y_{ij}-\mathbf{q}^\intercal (\mathbf{u_i}\circ \mathbf{v_j})}^2+\\
    &&& \sum_{i=1}^M \frac{\lambda_u}{2}\norm{\mathbf{u_i}}_2^2+\sum_{j=1}^N \frac{\lambda_v}{2}\norm{\mathbf{v_j}}_2^2+\frac{\lambda_p}{2}\norm{\mathbf{p}}_2^2+\frac{\lambda_q}{2}\norm{\mathbf{q}}_2^2,
\end{aligned}
\end{equation}
where $\Theta=\set{\mathbf{p},\mathbf{q}}$ are the parameters, and $\gamma$ is a trade-off of the online affinity $\mathbf{Y}$ for incorporating the user feedback. 

\paragraph{Offline Optimization}
When optimizing offline by using the historical ratings of all users, we use gradient descent with Adaptive Moment Estimation (Adam) optimizer ~\cite{kingma2014adam} for Eq.~\eqref{eq:log-post}, with $\gamma$ set to 0, since we do not have the historical question asking data and thus do not have $Y_{ij}$ for the question asking. Therefore, we do not train $\mathbf{q}$, instead set $\mathbf{q}$ to all-ones vector in this paper, but one can also train $\mathbf{q}$ using historical question asking data.
That is, the model variables $\mathbf{U}, \mathbf{V}, \mathbf{p}$ are learned by maximizing the log-posterior over the user and item latent vectors with fixed hyper-parameters, given the training observations $\mathbf{R}$. 

\paragraph{Online Optimization}
Since we aim to recommend items online, it is necessary to update the variables effectively and efficiently according to the user feedback. Thus, we optimize Eq.~\eqref{eq:log-post} by \ac{ALS} technique to update the model variables $\mathbf{u_i}$, and $\mathbf{v_j}$ in order to guarantee efficiency. Then we have our following derived closed-form solution:
{\small
\begin{align}
    & \mathbf{u_i} = \rbr{\mathbf{V_p}^\intercal \mathbf{V_p}+\gamma \mathbf{V_q}^\intercal \mathbf{V_q}+\lambda_u \mathbf{I}}^{-1}(\mathbf{V_p}^\intercal \mathbf{R}_i+\gamma \mathbf{V_q}^\intercal \mathbf{Y}_i)
    \label{equUV3}\\
    & \mathbf{v_j} = \rbr{\mathbf{U_p}^\intercal \mathbf{U_p}+\gamma \mathbf{U_q}^\intercal \mathbf{U_q}+\lambda_v \mathbf{I}}^{-1}(\mathbf{U_p}^\intercal \mathbf{R}_{.j}+\gamma \mathbf{U_q}^\intercal \mathbf{Y}_{.j})
\label{equUV4}
\end{align}
}%
where
\begin{displaymath}
\small
\begin{split}
    \mathbf{V_p}&=\mathbf{V} \diag{\mathbf{p}} , \\
    \mathbf{V_q}&=\mathbf{V} \diag{\mathbf{q}} , \\
    \mathbf{U_p}&=\mathbf{U} \diag{\mathbf{p}} , \\
    \mathbf{U_q}&=\mathbf{U} \diag{\mathbf{q}}.
\end{split}
\end{displaymath}

\noindent \ac{ALS} repeatedly optimizes one of $\mathbf{U}$ and $\mathbf{V}$ while temporarily fixing the other to be constant. After each question being asked and feedback received, we update $\mathbf{U}$ and $\mathbf{V}$. We assume that there is a target item related document $d^* \in \mathcal{D}$ and define an indicator vector $y^l_j$ for the $l$-th question, with each dimension $j$ corresponding to an item in the collection:
{\small
\begin{align}
y^l_j &= \mathbbm{1}\{e^{d_j}_l = e^{d^*}_l\}  \label{equUV5} , \\
Y_{j} &=\sum_{t = 0}^{l-1} y^t_j  \label{equUV6} ,
\end{align}
}%
where $e^{d_j}_l$ is true if the item related document ${d_j}$ contains the $l$-th requested entity $e_l$ (see details for the question construction in Section~\ref{sec:Questionasking}), and $\mathbbm{1}\{\cdot\}$ is an indicator function. $e^{d^*}_l$ expresses whether the target item contains the $l$-th requested entity $e_l$. This also represents the answer by the user, given that the user's answers are driven by a target item. Hence, for example if the question is ``Are you seeking for a [cotton] item?'' and the target item description includes ``cotton'' as an entity, then $y^l_j$ is 1 for all items that also have ``cotton'' as an important entity. If the question is ``Are you seeking for a [beach towel] item?'' and the target product does not contain a ``beach towel'' in its description or reviews (hence the answer of the user is ``no'') then $y^l_j$ is 1 for all the items that are not beach towels. $Y_j$ is the accumulated $y_j$ with the dimension corresponding to $j$-th item until the $l$-th question. 

Based on whether or not the target item is relevant to the requested entity, the feedback from user becomes a new or an updated observation for our system, and hence it is used to update $\mathbf{Y}$ related to the particular user, i.e.\ $\mathbf{Y}_i$, which is a vector of the online affinity for user $i$, with each of the dimension $Y_{ij}$ corresponding to $j$-th item. Then $\mathbf{u_i}$, and all item factors $\mathbf{V}$ are updated by Eq.~\eqref{equUV3} and Eq.~\eqref{equUV4}. Note that this observation only affects the current user's interaction session, and not any follow-up user interactions. 
As we ask about an entity $e$ and observe the user's response, the user's preference over the items which are consistent with the answer increases. The variance of the inferred noisy preferences over these items which is consistent with the answer as well as the variance of the nearby items in the learned embedding are reduced. The model's confidence in its belief over the user's preference on these items increases. As the system keeps asking questions to user $i$ and incorporates his/her responses, the latent user feature vectors $\mathbf{U}$ and latent item feature vectors $\mathbf{V}$ change and move towards the true underlying user and item latent vectors. 

After updating our matrix factorization model, we use the final user latent factor $\mathbf{U}$ and item latent factor $\mathbf{V}$ to computing $\mathbf{UV}^\top$ to yield a ranking of items to generate the recommendation list, which constitutes the recommendation module in Figure~\ref{fig:framework}. 

\subsection{Question Learning}
\label{sec:Questionlearning}
In this section, we describe how we select the next question to ask from the question pool (see Section~\ref{sec:Questionasking} for the question pool construction). 
After the offline initialization by using all of the historical ratings, the user initiates an interaction with our recommender system, our system asks a few questions to learn about the user latent factor, the item latent factor, and the user's belief. During this interactive phase, it is important to select the most informative questions that lead to learning effectively the user's preference, so as to minimize the number of questions asked and locate the target item effectively.

Similar to ~\citet{Wen:2013} and ~\citet{zou2018technology}, we use the estimated user preferences to help the question learning module to learn the most discriminative question to ask next. We model the user preferences for the items by a (multinomial) probability distribution $\pi^*$ over items $\mathcal{D}$, and the target item is drawn i.i.d. from this distribution. We also assume that there is a prior belief $\mathbb{P}$ over the user preferences $\pi^*$, which is a probability density function over all the possible realizations of $\pi^*$. 
\begin{equation}
\small
\mathbb{P}_{l} = Dir(\alpha +  \mathbf{Y}_i),
\end{equation}

\noindent where $\mathbb{P}$ is a Dirichlet distribution with parameter $\alpha$. Having applied the offline initialization of our matrix factorization model, items can be scored and ranked for each user, the rank of each item expresses our initial belief on the preference of items for each given user. This initial belief will be used to initialize the hyper-parameter $\alpha$ of the Dirichlet distribution. In particular, we set $\alpha_i$ for item $i$ to $1/(p_i+1)$, where $p_i$ is the index of item $i$ in the ranked list.  $\mathbf{Y}_i$ is the vector for the user $i$ with each dimension corresponding to accumulated $y^l_j$ until the $l$-th question.

Let $\mathbb{P}_l$ be the system’s belief over $\pi^*$ prior to the $l$-th question. 
We compute the user preferences $\pi^*_l(d)$ prior to the $l$-th question by:
\begin{equation}
\small
\pi^*_l(d) = \mathbb{E}_{\pi \sim \mathbb{P}_l} [\pi(d)]\forall d \in \mathcal{D} .
\label{equ2}
\end{equation}

The $\pi^*$ is a multinomial distribution over items $\mathcal{D}$, and $\mathbb{P}$ is modeled by the conjugate prior of the multinomial distribution, i.e.\ the Dirichlet distribution. From the properties of the Dirichlet distribution, the user preferences $\pi^*_l$ can be updated by counting and re-normalization of $\alpha$ and $\mathbf{Y}_i$. As the system keeps asking questions to the user and incorporates his/her response, the predicted belief and preferences about the user is updated accordingly. This belief tracker thus specifies the direction for moving towards the true underlying belief distribution and true user preferences. This predicted user preferences will be used for guiding the question selection. 

Same to ~\citet{Wen:2013} and ~\citet{zou2018technology}, we apply \ac{GBS} to find the entity that best splits the probability mass of predicted user preferences closest to two halves for the remaining of the items during the $l$-th question, as the nearly-optimal entity to ask. 
\begin{equation}
\small
e_l = \arg\min_{e}\Big|\sum_{d \in {\cal C_l}} (2 \mathbbm{1}\{e^d = 1\} - 1)\pi_l^*(d) \Big|
\label{equ4}
\end{equation}
where $e_l$ is the $l$-th chosen entity, ${\cal C_l}$ is the candidate version space containing the set of remaining items when asking the $l$-th question; the initial ${\cal C_l}$ is equal to $\mathcal{D}$,   $e^d$ expresses whether the item $d$ contains the entity $e$ or not. Specifically, for the entity embedding in this paper, the entity is represented by one-hot encoding, i.e.\ if the entity appears in a certain item documents, the value of the dimension corresponding to this item is 1 ($e^d=1$), otherwise the value of the dimension corresponding to this item is 0 ($e^d=0$). After each question is asked and the answer is obtained, the user preferences $\pi^*$ are updated by the belief tracker module. \ac{GBS} tend to select entities by minimizing the objective function of Eq.~\eqref{equ4}. This means, \ac{GBS} selects the entity which is able to split the sum of calculated user preferences corresponding to the item with $e^d=1$ and the sum of user preferences corresponding to the item with $e^d=0$ closest to two halves.

\subsection{Question Asking}
\label{sec:Questionasking}
The proposed method of learning informative questions to ask to users, 
depends on the availability of a pool of questions regarding informative terms. Given an item, the user should be able to answer questions about this item with a ``yes'' or a ``no'', having a reference to the relevant item (or item in mind).

In this work, we use the approach taken by ~\citet{zou2018technology}, and \citet{ zou2019learning} to extract meaningful short-phrases -- typically entities -- from the surface text to construct the question pool using the entity linking algorithm TAGME ~\cite{ferragina2010tagme}. These entities are recognized to comprise the most important characteristics of an item ~\cite{zou2018technology, zou2019learning}, and we generate questions about the presence or absence of these entities in the item related documents. One could also use other sources like labelled topics, extracted keywords, item categories and attributes, to construct questions.

In TAGME, each annotated short-phrase in unstructured text is weighted using a probability, 
that measures the reliability of that substring being a significant mention. 
Only the short-phrases with high probability should be considered as entities. In this paper, similar to ~\citet{ferragina2010tagme}, and after a set of preliminary experiments, we set the threshold to 0.1 and filter out the short-phrases whose probability is below 0.1. Prior to this, we also removed stop words such as ``about'', ``as well'' etc.. 

Having extracted the most important entities from the corpus, the proposed algorithm asks a sequence of questions in the form of ``Are you seeking for a [entity] related item?'' to locate the target item. In this case, the users can respond with a ``yes'', a ``no'' or a ``not sure'' according to their belief.

\begin{algorithm}[tb]
\small
\SetKwInOut{Input}{input}
\Input{A item document set, $\mathcal{D}$, the set of annotated entities in the documents, ${\cal E}$, the ratings $\mathbf{R}$, number of questions to be asked $N_q$
}
$l \leftarrow 0$\

$\mathbf{Y}_i \leftarrow \mathbf{0}$\

Offline intialization of our matrix factorization model: 
$\mathbf{U}, \mathbf{V} = QMF({\mathbf{R}}) $\

  ${Ranking}_l= Sort(\mathbf{U} {\mathbf{V}}^\top)$\
  
    $\alpha \leftarrow {Ranking}_l$\
  
  \While{$l < N_q$ and $| {\cal C_l}| > 1 $}{
  
  Compute the user belief with $\alpha$: $\mathbb{P}_l = Dir(\alpha + \mathbf{Y}_i)$ \ 
  
    Compute the user preferences with $\mathbb{P}_{l}(\pi)$: 
    $\pi^*_l(d) = \mathbb{E}_{\pi \sim \mathbb{P}_l}[\pi(d)] ~ \forall d \in \mathcal{D}$ \ 
    
    Find the optimal target entity by question learning: \
    
    $e_l = \arg\min_{e}\Big|\sum_{d \in {\cal C_l}} (2 \mathbbm{1}\{e^d = 1\} - 1)\pi_l^*(d) \Big|$\
    
    Ask the question about $e_l$, observe the reply $e^{d^*}_l$\ 
    
    Remove $e_l$ from question pool\
    
     ${\cal C_{l+1}} = {\cal C_l} \cap {d  \in \mathcal{D} : e^d_l=e^{d^*}_l}$\
    
    Update $\mathbf{Y}_i$ by the reply $e^{d^*}_l$ according to Eq.~\eqref{equUV5} and Eq.~\eqref{equUV6}\ 
    
    Update $\mathbf{U},\mathbf{V}$ by \ac{ALS} according to Eq.~\eqref{equUV3} and Eq.~\eqref{equUV4}\ 

    $l \leftarrow l + 1$\

  }
      
Generate recommendation list by updated $\mathbf{U},\mathbf{V}$:
$result = Sort(\mathbf{U}_{N_q}{\mathbf{V}_{N_q}}^\top)$
\captionsetup{font={small}}    
\caption{The proposed Qrec algorithm}
\label{algo1}
\end{algorithm}

\subsection{Question-based Recommender System}
\label{sec:Rec}
The algorithm of our question based recommender system is provided in Algorithm ~\ref{algo1}.  
Our Qrec model performs two rounds: the offline phase and the online phase. The offline phase includes \emph{line 3-5}, and the online phase includes \emph{line 6-17} in Algorithm ~\ref{algo1}. During the offline phase, we firstly initialize our model parameters offline by using the history rating data across all users. We make the assumption that we have access to historical user-item interaction data (e.g., rating or purchasing data), even though our system can work without it as well. When a new user session starts, we use the initialized user's latent factors and items' latent factors to yield the preliminary ranking of candidate items. We then utilize this ranking score to initialize the Dirichlet prior parameter $\alpha$. When there is a new user session starts in online phase, we calculate the user belief with this $\alpha$ and $\mathbf{Y}_i$. After that, we compute the user preferences with prior belief equal to $\mathbb{P}_l$, and find the optimal entity $e_l$ by \ac{GBS}. We ask whether the entity $e_l$ is present in the target item that the user wants to find, $d^*$, observe the reply $e^{d^*}_l$, remove $e_l$ from the question pool, and update the candidate version space ${\cal C_l}$. Then we update $\mathbf{Y}_i$ by the user response, and update the user latent factors $\mathbf{U}$ and the item latent factors $\mathbf{V}$ using \ac{ALS} based on the updated $\mathbf{Y}_i$. After the online question asking phase is over, the recommendation list is generated by sorting the inner product of the last updated user latent factors $\mathbf{U}_{N_q}$ and item latent factors $\mathbf{V}_{N_q}$.

\section{Experiments and Analysis}
\label{sec:exp}

\subsection{Experimental Setup}
\label{ExpSetup}
\subsubsection{Dataset.} 
In our experiments we use a collection of Amazon items \footnote{http://jmcauley.ucsd.edu/data/amazon/} ~\cite{mcauley2015image}. Each item contains rich metadata such as title, descriptions, categories, and reviews from users as well. Following ~\citet{van2016learning} and ~\citet{zou2019learning}, we use four different product domains from the Amazon product dataset, but due to the limited space, we only report two domains in this paper, which are "Home and Kitchen", and "Pet Supplies" respectively.
The documents associated with every item consist of the item description and the reviews provided by Amazon customers. On the two item domains, we use the same item list \footnote{Product list: https://github.com/cvangysel/SERT/blob/master/PRODUCT\_SEARCH.md} with ~\citet{van2016learning}, and filtered those items and users that appeared in less than five transactions to construct the user-item recommendation matrix like most of Collaborative Filtering papers ~\cite{he2017neural}. We randomly split the entire dataset of user-item interactions to a training, validation and testing set with 80\%, 10\% and 10\% split similar to other recommendation papers, e.g. ~\citet{sun2018conversational}. Statistics on the dataset are shown in Table ~\ref{table:data}. 

\begin{table}
\captionsetup{font={small}}
\caption{Statistics of the dataset. \#entity is the number of unique entities.}
\label{table:data}
\centering
\small
\begin{tabular}{p{0.18\columnwidth}{l}p{0.1\columnwidth}{l}p{0.1\columnwidth}{l}p{0.1\columnwidth}{l}p{0.1\columnwidth}{l}p{0.1\columnwidth}{l}}
\toprule
Dataset & \#users & \#items & \#ratings & density & \#entity\\
\hline 
Home and Kitchen & 9,124 & 557 & 10,108 & 0.20\% & 9,296\\
Pet Supplies & 2,248 & 2,475 & 15,488 & 0.28\% & 71,074\\
\bottomrule
\end{tabular}
\end{table}

\subsubsection{Parameter Setting} 
To learn the matrix factorization embedding, we set the hyper-parameters to the combination that achieved the highest pairwise accuracy in the offline observations: the maximum training iterations of PMF and our matrix factorization model is set to 100, and $\lambda_u = \lambda_v = \lambda_p = \lambda_q = 0.1$. The parameters $\gamma$, the dimension of the latent factors $K$, and the number of questions asked $N_q$ are decided in \RQRef{1}.

\subsubsection{Evaluation Metrics} 
We use average Recall at cut-off 5 (recall$@$5), Average Precision at 5 (AP$@$5), and Mean Reciprocal Rank (MRR) and Normalized Discounted Cumulative Gain (NDCG) as our evaluation metrics, which are commonly used metrics for capturing accuracy in recommendation ~\cite{christakopoulou2016towards, zhao2013interactive, zhang2018towards}. NDCG is calculated by top 100 items like other paper ~\cite{van2016learning}. 
The ground truth used to compute the aforementioned metrics is constructed by looking at the historical buying behavior of the user; an item is considered relevant if the user wrote a review and gave a rating to it, similar to other works ~\cite{zhang2018towards, van2016learning}.

\subsubsection{Baselines} 
 We compare our method with five baselines; the first two are static baselines, while the other three are interactive baselines. In particular the baselines are: (1) \textbf{PMF}, which is a typical, static recommendation approach; 
(2) \textbf{NeuMF} ~\cite{he2017neural}, which is one of the state of the art approaches of collaborative filtering and widely used as the baseline by other papers. 
 (3) \textbf{QMF+Random}, which uses our proposed matrix factorization for offline initialization and then randomly chooses a question from the question pool to ask; (4) \textbf{SBS}, which is the sequential Bayesian search algorithm. We applied the SBS ~\cite{Wen:2013} to our recommendation task and uses the same question asking strategy with our Qrec model, but with the uniform prior; 
 and (5) \textbf{PMMN} ~\cite{zhang2018towards}, the Personalized Multi-Memory Network model, which is a state-of-the-art conversational recommender system asking questions on aspect-value pairs. 
For the PMF, QMF+Random, and SBS baselines, we use the same parameter setting with our Qrec model. For the NeuMF and PMMN, we use the optimal parameters reported in the corresponding paper and tuned their hyper-parameters in the same way as they reported.
 
\subsubsection{Simulating Users} 
Our experimentation depends on users responding to questions asked by our method. In this paper we follow recent work ~\cite{zhang2018towards, sun2018conversational, zou2018technology, zou2020towards, zou2019learning} and simulate users. We also conduct a small user study described next. During the simulation, we follow the approach proposed by ~\citet{zou2018technology} and ~\citet{zou2019learning}, i.e.\ we assume that the user will respond to the questions with full knowledge of whether an entity is present or not in the target item. Hence, we assume that the user will respond with ``yes'' if an entity is contained in the target item documents and ``no'' if an entity is absent. This simulation also follows the one used by ~\citet{zhang2018towards}, which assumes that the user has perfect knowledge of the value of an aspect for the target product. 

\subsubsection{Online User Study}
To confirm some of the assumptions made in this work and test how well our recommender system works ``in-situ'' we also conduct a small online user study. The ideal users would be ones who have actually bought a number of items on an online shopping platform and now converse with our system embedded in the platform to find their next target item. In the absence of such a user base and commercial recommender system we use a crowdsourcing platform. First, we let the crowd worker select a product category she feels familiar with. Then, we randomly sample a product from our test data as a target product. To let the user familiarize herself with the target product we provide her with a product image, title, description, and the entities extracted from the product reviews. After the user indicates that she is familiar with the product and the conversation with the system can start, the information of the target item disappears from the screen and a question is selected by our algorithm to be asked to the user. The user needs to provide an answer to the question according to the information she read in the previous step, and then our system updates according to the user answer. With each question being answered, the user is shown a grid (4-by-4) of the pictures of sixteen top ranked items. The user can stop answering questions any time during her interaction with the system. When stoping the interaction with the system, users are asked a number of exit questions about their experiences with the system.

\begin{figure}
\captionsetup{font={small}}
\includegraphics[width=0.47\columnwidth]{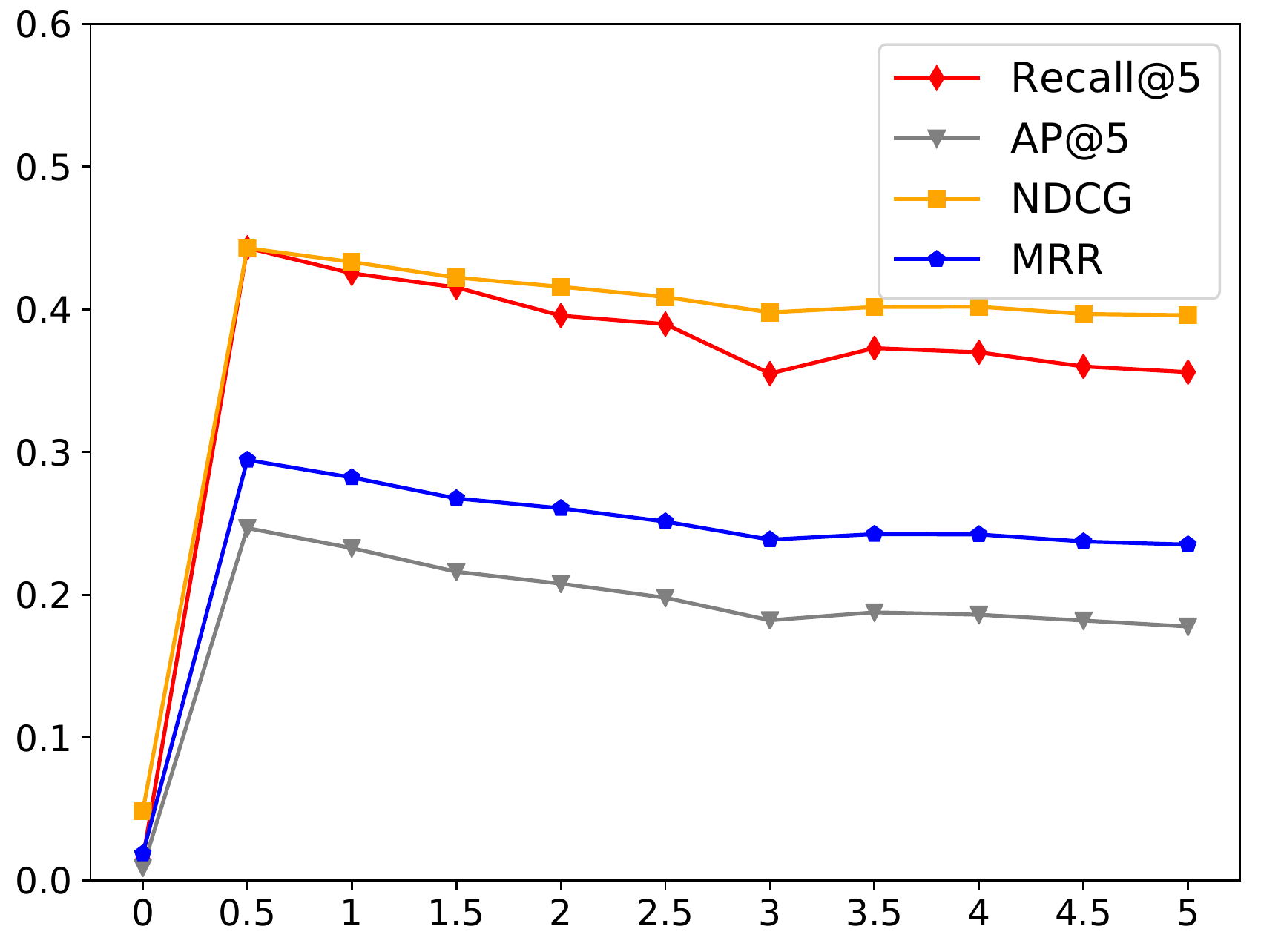} 
  \includegraphics[width=0.48\columnwidth]{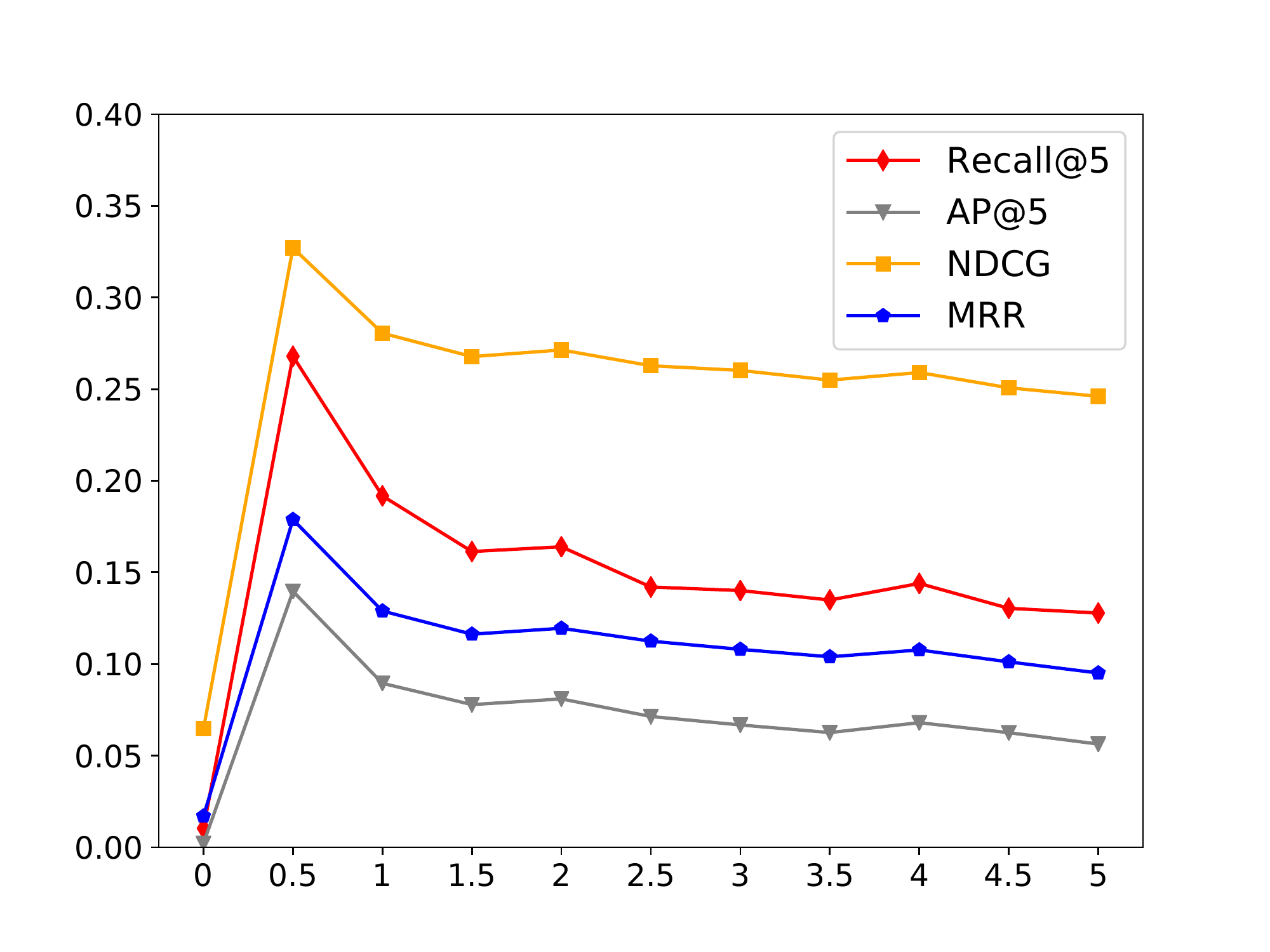} 
  \includegraphics[width=0.47\columnwidth]{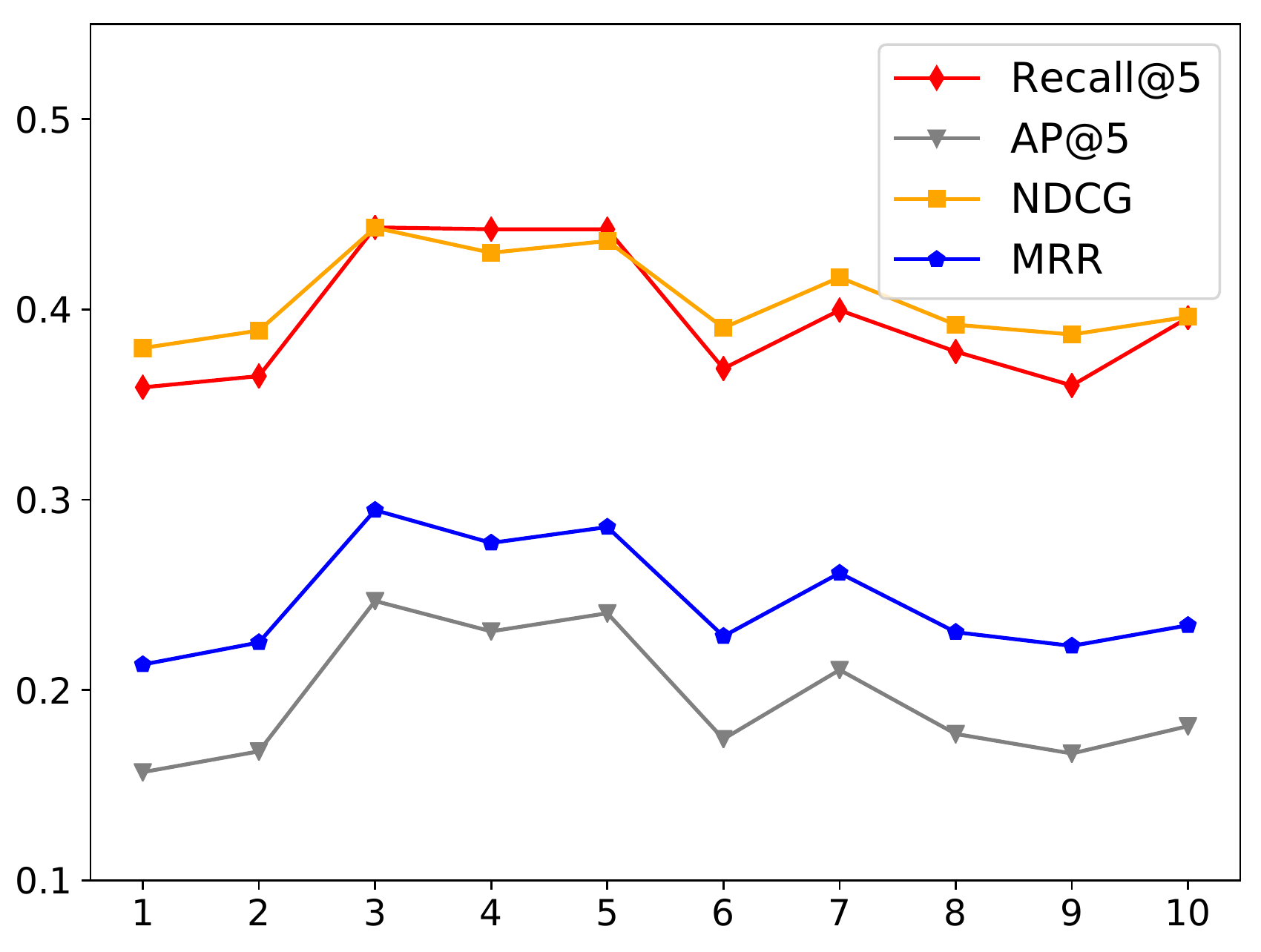} 
  \includegraphics[width=0.48\columnwidth]{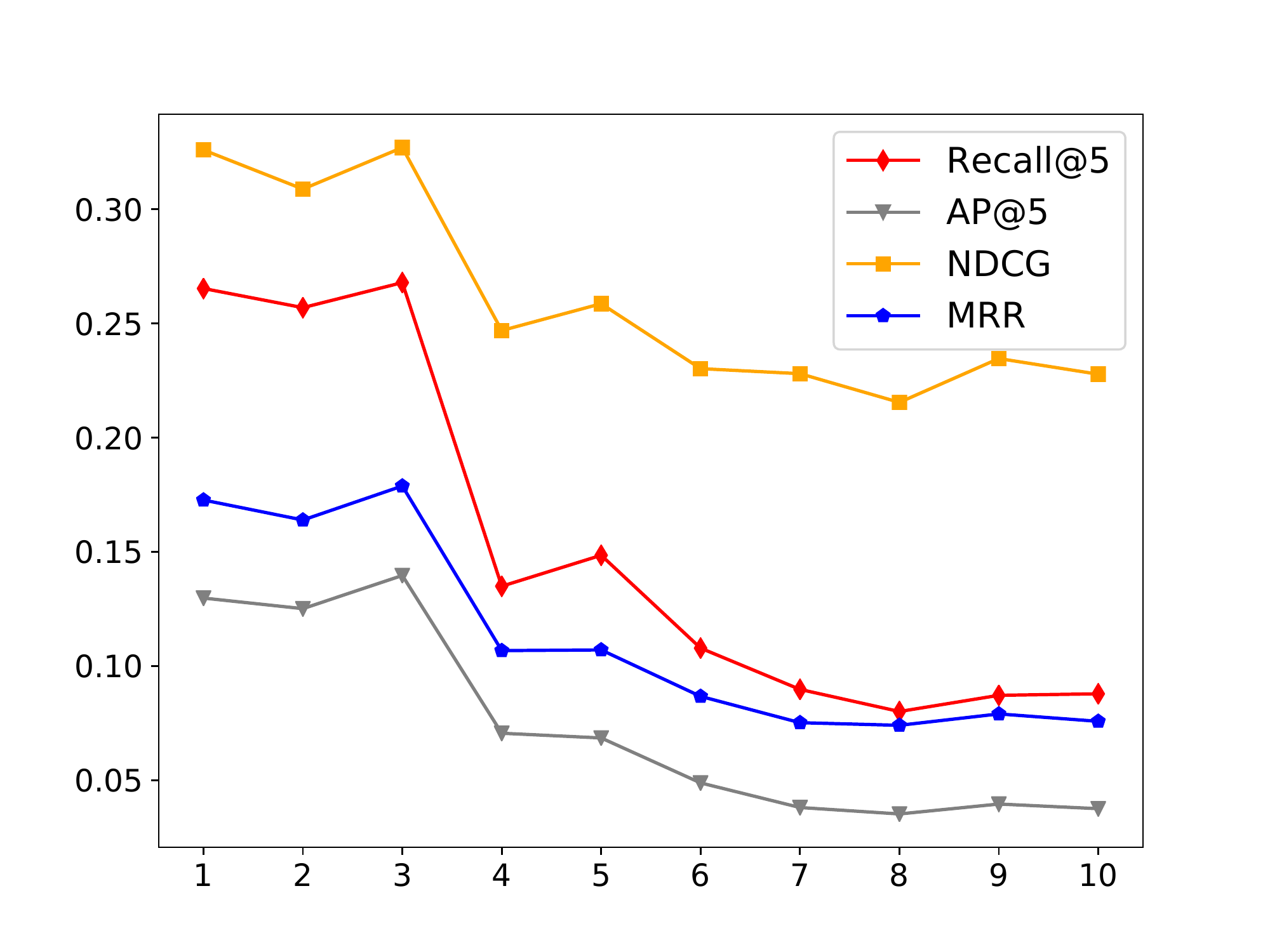}
    \caption{The impact of the trade-off parameter $\gamma$ (top), and the dimension of the latent factors $K$ (bottom) on "Home and Kitchen" (left) and "Pet Supplies" (right) categories.}
  \label{fig:para}
\end{figure}

\subsubsection{Research Questions.} 
Through the experiments in this work we aim to answer the following research questions:
\RQ{1}{What is the impact of the trade-off $\gamma$, the dimension of the latent factors $K$, and the number of questions asked $N_q$?}
\RQ{2}{How effective is Qrec compared to prior works?}
\RQ{3}{How effective is Qrec for the cold-start user and the cold-start item problem?}
\RQ{4}{Does the offline initialization help?}
\RQ{5}{Are the assumptions made in this work along with the effectiveness of our algorithm confirmed by a user study?}

\begin{table*}
\captionsetup{font={small}}
\caption{The comparison with PMF, NeuMF, QMF+Random, SBS, and PMMN on the "Home and Kitchen" (top) and the "Pet Supplies" (bottom) categories.  \#. represents the number of asked questions. QMF+Rand. represents the QMF+Randam model. Our proposed model achieve highest results when compared with interactive baselines, and our model performs better than the state of the art collaborative filtering model NeuMF on all of four different metrics with less than 5 questions.
}
\label{table:3}
\centering
  \small
\begin{tabular}{c|ccccccc|ccccccc}
\toprule
& \multicolumn{7}{p{0.95\columnwidth}<{\centering}|} {recall$@$ 5} & \multicolumn{7}{p{0.95\columnwidth}<{\centering}} {AP$@$ 5}\\
\hline
 \#. & PMF & QMF & NeuMF & QMF+Rand. & SBS & PMMN & Qrec & PMF & QMF & NeuMF & QMF+Rand. & SBS & PMMN  & Qrec \\
 \hline
2 & 0.011 & 0.062 & \textbf{0.222} & 0.075 & 0.060 & 0.073 & 0.130 & 0.004 & 0.037 & \textbf{0.121} & 0.047 & 0.028 & 0.021 & 0.072 \\
5 & 0.011 & 0.062 & 0.222 & 0.095 & 0.353 & 0.194 & \textbf{0.443} & 0.004 & 0.037 & 0.121 & 0.064 & 0.170 & 0.091 & \textbf{0.247} \\
10 & 0.011 & 0.062 & 0.222 & 0.121 & 0.883 & 0.216 & \textbf{0.943} & 0.004 & 0.037 & 0.121 & 0.088 & 0.661 & 0.105 & \textbf{0.884} \\
15 & 0.011 & 0.062 & 0.222 & 0.151 & 0.933 & 0.216 & \textbf{0.982} & 0.004 & 0.037 & 0.121 & 0.117 & 0.863 & 0.105 & \textbf{0.973} \\
20 & 0.011 & 0.062 & 0.222 & 0.188 & 0.962 & 0.216 & \textbf{0.995} & 0.004 & 0.037 & 0.121 & 0.144 & 0.911 & 0.105 & \textbf{0.990} \\
\hline
& \multicolumn{7}{p{0.95\columnwidth}<{\centering}|} {NDCG} & \multicolumn{7}{p{0.95\columnwidth}<{\centering}} {MRR}\\
\hline
 \#. & PMF & QMF & NeuMF & QMF+Rand. & SBS & PMMN & Qrec & PMF & QMF & NeuMF & QMF+Rand. & SBS & PMMN  & Qrec \\
 \hline
2 & 0.036 & 0.082 & 0.183 & 0.113 & 0.181 & 0.212 & \textbf{0.215} & 0.011 & 0.048 & \textbf{0.136} & 0.062 & 0.053 & 0.050 & 0.100 \\
5 & 0.036 & 0.082 & 0.183 & 0.131 & 0.389 & 0.300 & \textbf{0.443} & 0.011 & 0.048 & 0.136 & 0.079 & 0.226 & 0.135 & \textbf{0.295} \\
10 & 0.036 & 0.082 & 0.183 & 0.158 & 0.749 & 0.310 & \textbf{0.915} & 0.011 & 0.048 & 0.136 & 0.104 & 0.671 & 0.147 & \textbf{0.889} \\
15 & 0.036 & 0.082 & 0.183 & 0.184 & 0.899 & 0.310 & \textbf{0.980} & 0.011 & 0.048 & 0.136 & 0.132 & 0.869 & 0.147 & \textbf{0.975} \\
20 & 0.036 & 0.082 & 0.183 & 0.211 & 0.935 & 0.310 & \textbf{0.993} & 0.011 & 0.048 & 0.136 & 0.159 & 0.915 & 0.147 & \textbf{0.991} \\
 \hline
\end{tabular}
\BlankLine
\begin{tabular}{c|ccccccc|ccccccc}
 \hline
& \multicolumn{7}{p{0.95\columnwidth}<{\centering}|} {recall$@$ 5} & \multicolumn{7}{p{0.95\columnwidth}<{\centering}} {AP$@$ 5}\\
\hline
 \#. & PMF & QMF & NeuMF & QMF+Rand. & SBS & PMMN & Qrec & PMF & QMF & NeuMF & QMF+Rand. & SBS & PMMN  & Qrec \\
 \hline
2 & 0.008 & 0.016 & \textbf{0.214} & 0.016 & 0.007 & 0.056 & 0.076 & 0.005 & 0.008 & \textbf{0.119} & 0.008 & 0.003 & 0.026 & 0.030 \\
5 & 0.008 & 0.016 & 0.214 & 0.017 & 0.052 & 0.139 & \textbf{0.268} & 0.005 & 0.008 & 0.119 & 0.009 & 0.024 & 0.095 & \textbf{0.140} \\
10 & 0.008 & 0.016 & 0.214 & 0.033 & 0.668 & 0.143 & \textbf{0.966} & 0.005 & 0.008 & 0.119 & 0.024 & 0.393 & 0.097 & \textbf{0.770} \\
15 & 0.008 & 0.016 & 0.214 & 0.035 & 0.952 & 0.143 & \textbf{0.999} & 0.005 & 0.008 & 0.119 & 0.025 & 0.823 & 0.098 & \textbf{0.997} \\
20 & 0.008 & 0.016 & 0.214 & 0.039 & 0.994 & 0.143 & \textbf{1.000} & 0.005 & 0.008 & 0.119 & 0.029 & 0.961 & 0.098 & \textbf{1.000} \\
\hline
& \multicolumn{7}{p{0.95\columnwidth}<{\centering}|} {NDCG} & \multicolumn{7}{p{0.95\columnwidth}<{\centering}} {MRR}\\
\hline
 \#. & PMF & QMF & NeuMF & QMF+Rand. & SBS & PMMN & Qrec & PMF & QMF & NeuMF & QMF+Rand. & SBS & PMMN  & Qrec \\
 \hline
2 & 0.012 & 0.056 & \textbf{0.179} & 0.069 & 0.032 & 0.121 & 0.141 & 0.007 & 0.019 & \textbf{0.134} & 0.021 & 0.010 & 0.046 & 0.054 \\
5 & 0.012 & 0.056 & 0.179 & 0.068 & 0.190 & 0.231 & \textbf{0.327} & 0.007 & 0.019 & 0.134 & 0.022 & 0.054 & 0.115 & \textbf{0.179} \\
10 & 0.012 & 0.056 & 0.179 & 0.038 & 0.557 & 0.233 & \textbf{0.830} & 0.007 & 0.019 & 0.134 & 0.028 & 0.427 & 0.117 & \textbf{0.774} \\
15 & 0.012 & 0.056 & 0.179 & 0.039 & 0.870 & 0.233 & \textbf{0.998} & 0.007 & 0.019 & 0.134 & 0.028 & 0.829 & 0.117 & \textbf{0.997} \\
20 & 0.012 & 0.056 & 0.179 & 0.049 & 0.971 & 0.233 & \textbf{1.000} & 0.007 & 0.019 & 0.134 & 0.034 & 0.961 & 0.117 & \textbf{1.000}\\ 
\bottomrule
\end{tabular}
\end{table*}

\subsection{Impact of Parameters (\RQRef{1})}
In \RQRef{1}, we examine the impact of the trade-off parameter $\gamma$, the dimension of the latent factors $K$, and the number of questions asked $N_q$ over the effectiveness of our model. We compare the performance for different parameters. When the comparison for the given parameter, we fix the other two parameters. 
The performance evolution of different $\gamma$ and different dimension of the latent factors $K$ on the two categories is shown in Figure ~\ref{fig:para}, and the results of different number of questions on the two categories can be seen in "Qrec" column of Table ~\ref{table:3}. The $\gamma$ ranges from 0 to 5 with a step of 0.5, and the $K$ ranges from 1 to 10 with a step of 1.
As one can observe, with the increase of $\gamma$, the performance first improves and then drops. The best $\gamma$ is 0.5 on the two categories. $\gamma$ can control how much online user feedback is incorporated into the user latent factor and item latent factor. In particular, when $\gamma$ is 0, i.e.\ the online updating do not take the user feedback (i.e.\ $\mathbf{Y}$) into account, as expected the performance is very bad. 
As for the dimension of the latent factors $K$, the overall performance trend also rises and then goes down with the increase of $K$. This suggests that the dimension of the latent factors $K$ should not be set too high or too low. In this paper, we set it to the optimal value, i.e.\ 3. 
Unless mentioned otherwise, in the rest of research questions, we use the optimal parameter $\gamma$, which is 0.5, and $K$ we used is the optimal value 3. 
To figure out the impact of the number of asked questions, we vary $N_q$ and see the performance shown in "Qrec" column of Table ~\ref{table:3}. As shown in Table ~\ref{table:3}, the performance of our Qrec model increases on all metrics with the increase of the number of questions, as expected. The more questions asked, the better the user needs are captured, and the closer the modeled user latent factor and item latent factor are to the true real-time user and item latent factors. Furthermore, the performance of Qrec reaches very good performance already, within the first 10 questions, while asking more than 10 questions does not add much regarding the performance.

\subsection{Performance Comparison (\RQRef{2})}
To answer how effective is our proposed method compared to prior works, we compare our results with five baselines, PMF, NeuMF, QMF+Random, SBS, and PMMN. 
The results on the two categories are shown in Table  ~\ref{table:3}. From Table ~\ref{table:3}, we can see that our proposed model, Qrec, achieves the highest results on all four metrics compared with the interactive baselines QMF+Random, SBS, and PMMN, on these two categories, which suggests that our question-based recommender system Qrec is effective. Our Qrec model performs better than QMF+Random, this suggests that our question selection is effective. There are few fluctuations on some metrics for QMF+Random with different number of questions asked, this is because the uncertainty of random question selection in different number of questions asked. Our Qrec model is superior to the SBS model, this suggests that using the prior from the offline initialization is beneficial. We will further discuss this in \RQRef{4}. Further, our Qrec model performs better than PMMN ~\cite{zhang2018towards}, especially after 5 questions asked. This might be explained by the fact that asking questions on extracted entities can gather more information from users and is able to better learn user true preferences than asking questions on aspect-value pairs. Further, what we indeed observed is that the results of all four metrics regarding PMMN do not increase much and the result differences between PMMN and our Qrec become big when the number of questions is larger than 10. The reason for this is the fact that it is rather difficult to extract more than 10 aspect-value pairs from each user review for a certain item. As a consequence, there are no more available questions to ask, and thus the metric results never increase. Overall, this suggests that asking question on extracted entities is more effective. 

It also can be observed that our proposed matrix factorization model achieves better performance than PMF on the four metrics, this suggests that our proposed matrix factorization model is rather helpful. The reason might be because that adding the parameter $P$ improves the model capability of fitting. The NeuMF model outperforms linear models PMF and QMF, this is because the nonlinear deep neural model can obtain more subtle and better latent representations. But note that the stacked neural network structures also make them difficult to train and incur a high computational cost.
Specifically, our model is able to achieve better results than the NeuMF model on all of four different metrics with less than 5 questions. With more questions being asked, the result differences between NeuMF and our Qrec become bigger. This shows that interactive or question-based recommendation can improve the performance over static models as interactive or question-based recommendation can continuously learn from the user. 

\begin{table}
\captionsetup{font={small}}
\caption{The results on cold-start tuples. The top table represents the cold-start user tuples on "Home and Kitchen" and bottom table represents the cold-start item tuples on "Pet Supplies" category. Our Qrec model can still achieve high performance for cold start users and cold start items.
}
\label{table:4}
\centering
  \small
\begin{tabular}{c|cccc}
\toprule
\# of questions & recall$@$5 & AP$@$5 & NDCG & MRR\\
 \hline
PMF & 0.005 & 0.002 & 0.039 & 0.011 \\
2 & 0.127 & 0.071 & 0.215 & 0.099 \\
5 & 0.448 & 0.245 & 0.442 & 0.293 \\
10 & 0.944 & 0.883 & 0.914 & 0.889 \\
15 & 0.985 & 0.974 & 0.981 & 0.976 \\
20 & 0.996 & 0.991 & 0.994 & 0.992 \\
\hline 
\end{tabular}
\BlankLine
\begin{tabular}{c|cccc}
\hline
\# of questions & recall$@$5 & AP$@$5 & NDCG & MRR\\
\hline
PMF & 0.000 & 0.000 & 0.000 & 0.000 \\
2 & 0.000 & 0.000 & 0.008 & 0.003 \\
5 & 0.046 & 0.011 & 0.157 & 0.035 \\
10 & 0.853 & 0.561 & 0.676 & 0.576 \\
15 & 0.991 & 0.961 & 0.971 & 0.962 \\
20 & 1.000 & 1.000 & 1.000 & 1.000 \\
\bottomrule
\end{tabular}
\end{table}

\begin{table*}
\captionsetup{font={small}}
\caption{Results for the effects of offline initialization on the "Home and Kitchen" (top) and the "Pet Supplies" (bottom) categories. Qrec\_offl. represents the Qrec including offline initialization, Qrec\_rand. represents the Qrec with random initialization (i.e, excluding offline initialization). The Qrec including offline initialization is superior to the Qrec excluding offline initialization.
}
\label{table:init}
\centering
  \small
\begin{tabular}{c|cc|cc|cc|cc}
\toprule
& \multicolumn{2}{p{0.4\columnwidth}<{\centering}|}{recall$@$5} & \multicolumn{2}{p{0.4\columnwidth}<{\centering}|}{AP$@$5} & \multicolumn{2}{p{0.4\columnwidth}<{\centering}|}{NDCG} & \multicolumn{2}{p{0.4\columnwidth}<{\centering}}{MRR}\\
\hline 
\# of questions & Qrec\_offl.& Qrec\_rand. & Qrec\_offl.& Qrec\_rand. & Qrec\_offl.& Qrec\_rand. & Qrec\_offl.& Qrec\_rand. \\
 \hline
 2 & 0.13 & 0.08 & 0.07 & 0.04 & 0.22 & 0.19 & 0.10 & 0.07 \\
5 & 0.44 & 0.34 & 0.25 & 0.17 & 0.44 & 0.39 & 0.29 & 0.22 \\
10 & 0.94 & 0.93 & 0.88 & 0.88 & 0.91 & 0.91 & 0.89 & 0.89 \\
15 & 0.98 & 0.98 & 0.97 & 0.97 & 0.98 & 0.98 & 0.97 & 0.97 \\
20 & 1.00 & 0.99 & 0.99 & 0.99 & 0.99 & 0.99 & 0.99 & 0.99 \\
\hline
\end{tabular}
\BlankLine
\begin{tabular}{c|cc|cc|cc|cc}
\hline
& \multicolumn{2}{p{0.4\columnwidth}<{\centering}|}{recall$@$5} & \multicolumn{2}{p{0.4\columnwidth}<{\centering}|}{AP$@$5} & \multicolumn{2}{p{0.4\columnwidth}<{\centering}|}{NDCG} & \multicolumn{2}{p{0.4\columnwidth}<{\centering}}{MRR}\\
\hline 
\# of questions & Qrec\_offl.& Qrec\_rand. & Qrec\_offl.& Qrec\_rand. & Qrec\_offl.& Qrec\_rand. & Qrec\_offl.& Qrec\_rand. \\
\hline
2 & 0.08 & 0.02 & 0.03 & 0.01 & 0.14 & 0.05 & 0.05 & 0.02 \\
5 & 0.27 & 0.11 & 0.14 & 0.06 & 0.33 & 0.24 & 0.18 & 0.09 \\
10 & 0.97 & 0.97 & 0.77 & 0.65 & 0.83 & 0.74 & 0.77 & 0.65 \\
15 & 1.00 & 1.00 & 1.00 & 1.00 & 1.00 & 1.00 & 1.00 & 1.00 \\
20 & 1.00 & 1.00 & 1.00 & 1.00 & 1.00 & 1.00 & 1.00 & 1.00 \\
\bottomrule
\end{tabular}
\end{table*}

\subsection{Cold Start Performance Analysis (\RQRef{3})}

To explore if our proposed method is effective for the cold-start user and the cold-start item problem or not, we extract cold-start user tuples (i.e.\ user-item interactions in which the user never appear in the training set) and cold-start item tuples (i.e.\ user-item interactions in which the item never appear in the training set) from our testing dataset. Because there are very few cold-start item tuples in "Home and Kitchen" category, and very few cold-start user tuples in  "Pet Supplies" category, to the extent that results would not be reliable, we only use cold-start user tuples from the "Home and Kitchen" category and cold-start item tuples from the "Pet Supplies" category to validate the cold-start addressing ability of our model.
Statistics on two categories shows that there are about 84\% cold-start user tuples on the "Home and Kitchen" category and about 7\% cold-start item tuples on the "Pet Supplies" category. The results on the two categories are shown in Table ~\ref{table:4}. As it is observed, our Qrec model can still achieve high recall$@$5, AP$@$5, NDCG, and MRR for cold start users and cold start items. As it is known, PMF does not really work for cold start users and cold start items, which is indeed what we observe. We conclude that our Qrec model is capable of tackling the cold-start recommendation problem. 

\subsection{Contribution of Offline Initialization (\RQRef{4})}
In this research question, we investigate the effect of our offline initialization. We compare the performance results including the offline initialization and the performance results excluding the offline initialization of our model (i.e.\ random initialization for the model parameters when the new user session starts). Our hypothesis is that the offline learned parameters from the historical ratings capture some general trend and provide a generic prior to guide the model. Indeed, the results shown in Table ~\ref{table:init} demonstrates the model with offline initialization achieves higher performance than the one without offline initialization, especially when the early stage of question asking (here: the number of asked questions is less than 10). Based on the observation of performance improvements when initializing the model from the offline data, we conclude that using offline initialization is highly beneficial.

\subsection{Online User Study (\RQRef{5})}
\label{sec:userstudy}

In this research question we first want to examine the assumptions made in this work. In particular, we first want to understand how many questions actual users are willing to answer, how well do they answer them, and how is their experience with the system. 
We collected 489 conversations made between our system and 21 crowd workers on 33 target items. 
From the collected data, we observe that users answered an average number of 15 questions per target item in the system (with the median being 12). Further, in the exit questionnaire, 71.4\% of the users declare that they are willing to answer between 10 and 20 questions. Despite a median time of 5 seconds to answer a question, in the exit questionnaire, 95.2\% of the users indicate that the system's questions were easy to answer. From the results we collected, most of the users think the conversational system is helpful and they will use it in the future. In particular, 81\% of users found the experience positive, 14.3\% neutral, and 4.7\% negative. Last but not least, the users provided the correct answers to the system's question 95\% of the time, they were not sure about their answers 3.5\% of the time, and they gave the wrong answers (i.e.\ their answers disagreed with the description of the product) 1.5\% of the time.

The second important question is how well the system performed. We measured performance after 5, 10, 15, and 20 queries asked (for those conversations that had this number of questions), as well as the performance when the user indicated that she wanted to stop. The results are shown in Table~\ref{table:study}, and are in agreement with the Qrec results of Table~\ref{table:3}.
 
\section{Conclusions and Discussion}
\label{sec:conc}
In this paper, we propose a novel question-based recommendation method, Qrec, which directly queries users on the automatically extracted entities in relevant documents. Our model is initialized offline by our proposed matrix factorization model QMF and updates the user and item latent factors online by incorporating the modeling of the user answer for the selected question. Meanwhile, our model tracks the user belief and learns a policy to select the best question sequence to ask. Experiments on the Amazon product dataset demonstrate that the effectiveness of the Qrec model compared to existing baselines.

In this work, the questions asked to users are based on the presence or absence of entities in the target items, following past work. Richer type of questions could be constructed by using other sources such as categories, keywords, labelled topics ~\cite{zou2015non, zou2017towards}, structural item properties, and domain-specific informative terms. Also, we ignore the fact that entities may be semantically related to the target item even though they are not contained lexically in the item documents.
Further, we leave the number of questions asked as a parameter to be predefined instead of algorithmically decided. 
Our work uses a stand-alone algorithm that learns the informativeness of questions to ask based on \ac{GBS}. One can also use other techniques (e.g., reinforcement learning) to learn the optimal question asking strategy, or incorporate more factors, e.g., the relatedness and importance level of different informative terms, to extend the work.
Still, the user may change their target item during the interaction with the system ~\cite{plua2002collaborative}. Theoretically our method is able to deal with this kind of situation, with new answers received gradually for the new target item. 
Last, we conduct a small user study, however a larger and in-situ user study by intervening at the interface of a commercial recommender system would be more informative. We leave all these as future work.

\begin{table}
\captionsetup{font={small},aboveskip=3pt}
  \caption{System effectiveness on user study. Results are in agreement with the Qrec results of Table~\ref{table:3}.}
  \centering
  \small
  \begin{tabular}{c|cccc}
    \toprule
    \# of questions & recall$@$5 & AP$@$5 & NDCG & MRR\\
    \hline
    5&   0.333&  0.082& 0.305& 0.129\\
    10&  0.848& 0.717& 0.777& 0.727\\
    15&  0.879& 0.760& 0.806& 0.762\\
    20&  0.909& 0.775& 0.820& 0.776\\
    stopping & 0.939 & 0.790 & 0.834 & 0.791\\
    \bottomrule
  \end{tabular}
  \label{table:study}
\end{table}

\bibliographystyle{ACM-Reference-Format}
\bibliography{bibfile} 

\end{document}